%% file: main.tex
\documentclass[journal]{IEEEtran}
\usepackage{cite}
\usepackage{amsmath,amssymb,amsfonts}
\usepackage{algorithmic}
\usepackage{graphicx}
\usepackage{textcomp}
\usepackage{xcolor}
\usepackage[hyphens]{url}
\usepackage{subfig}
\usepackage{array}
\usepackage{makecell}
\usepackage{cleveref}
\usepackage{floatrow}
\def\BibTeX{{\rm B\kern-.05em{\sc i\kern-.025em b}\kern-.08em
   T\kern-.1667em\lower.7ex\hbox{E}\kern-.125emX}}

\usepackage{xcolor}
\usepackage{soul}

\newcommand{\hlc}[2][yellow]{{%
    \colorlet{foo}{#1}%
    \sethlcolor{foo}\hl{#2}}%
}
\renewcommand{\hlc}[2][yellow]{#2}
\title{Spatially parallel decoding for multi-qubit lattice surgery}

\author
{Sophia Fuhui Lin\textsuperscript{*}$^{1,2}$, Eric C. Peterson$^{1}$, Krishanu Sankar$^{1}$, Prasahnt Sivarajah$^{1}$\\
\normalsize{$^{1}$AWS Center for Quantum Computing}\\
\normalsize{$^{2}$University of Chicago}
\thanks{\textsuperscript{*}Corresponding author: Sophia Lin (email: lfsophia@amazon.com).}
}

\begin{document}
\makeatletter 
  \@namedef{figure}{\killfloatstyle\def\@captype{figure}\FR@redefs
    \flrow@setlist{{figure}}%
    \columnwidth\columnwidth\edef\FBB@wd{\the\columnwidth}%
    \FRifFBOX\@@setframe\relax\@@FStrue\@float{figure}}%
\makeatother

\maketitle
\thispagestyle{plain}
\pagestyle{plain}

\input{sections/abstract}

\input{sections/introduction}

\input{sections/background}

\input{sections/related_work}

\input{sections/parallel_windows}

\input{sections/logical_err}

\input{sections/throughput}

\input{sections/conclusion}

\section*{Acknowledgements}
This work is funded in part by the STAQ project under award NSF Phy-1818914/232580 and in part by the NSF Quantum Leap Challenge Institute for Hybrid Quantum Architectures and Networks (NSF Award 2016136).


\bibliographystyle{IEEEtranS}
\bibliography{refs}

\end{document}

%% file: sections/abstract.tex
\begin{abstract}
Running quantum algorithms protected by quantum error correction requires a real time, classical decoder. To prevent the accumulation of a backlog, this decoder must process syndromes from the quantum device at a faster rate than they are generated. Most prior work on real time decoding has focused on an isolated logical qubit encoded in the surface code. However, for surface code, quantum programs of utility will require multi-qubit interactions performed via lattice surgery. A large merged patch can arise during lattice surgery --- possibly as large as the entire device. This puts a significant strain on a real time decoder, which must decode errors on this merged patch and maintain the level of fault-tolerance that it achieves on isolated logical qubits. 

These requirements are relaxed by using spatially parallel decoding, which can be accomplished by dividing the physical qubits on the device into multiple overlapping groups and assigning a decoder module to each. We refer to this approach as \textit{spatially parallel windows}. While previous work has explored similar ideas, none have addressed system-specific considerations pertinent to the task or the constraints from using hardware accelerators. In this work, we demonstrate how to configure spatially parallel windows, so that the scheme (1) is compatible with hardware accelerators, (2) supports general lattice surgery operations, (3) maintains the fidelity of the logical qubits, and (4) meets the throughput requirement for real time decoding. Furthermore, our results reveal the importance of optimally choosing the buffer width to achieve a balance between accuracy and throughput --- a decision that should be influenced by the device's physical noise.
\end{abstract}

%% file: sections/introduction.tex
\section{Introduction}
Given the error rates experienced by quantum computers, quantum error correction (QEC) is necessary for running large-scale quantum applications. QEC protects quantum information by encoding each logical qubit in multiple physical \textit{data} qubits, and using another set of physical \textit{syndrome} qubits to detect errors on the data qubits. In each measurement cycle, syndrome qubits extract parity information from the data qubits on which they act. This parity information is then sent to a classical decoder that decodes the syndromes and reports a correction. 

Here we focus on the surface code~\cite{dennis2002topological,fowler2012surface,Kitaev03}, a popular QEC that tolerates relatively high physical noise, supports relatively easy logical operations, and only requires nearest-neighbor grid connectivity. In recent years, academic and industry labs have experimentally demonstrated small instances of a surface code logical memory~\cite{google2023suppressing,krinner2022realizing,zhao2022realization} and lattice surgery~\cite{erhard2021entangling}. While an offline decoder --- applied after the experiment has completed --- is sufficient for such demonstrations, applications with nontrivial information processing will require real time decoding~\cite{fowler2012surface}. 


A requirement for real time decoding is that the throughput of the decoder match the rate of syndrome measurements, which avoids an exponential backlog of data~\cite{terhal2015quantum}. And the requirement can be quite strict: each syndrome measurement cycle on a superconducting device can be completed in $\sim 1$ $\mu s$ (921 ns in~\cite{google2023suppressing}). The strict timescales have persuaded the community to explore hardware accelerators for the task, such as Field Programmable Gate Arrays (FPGAs)~\cite{das2022lilliput,vittal2023astrea,liyanage2023scalable}, Application Specific Integrated Circuits (ASICs)~\cite{das2022afs}, or on-chip SFQ-based superconducting digital circuits~\cite{ravi2023better,holmes2020nisq+,ueno2021qecool}. While their software counterparts — designed to run on general-purpose CPUs~\cite{higgott2023sparse} — offer more flexibility, hardware decoders promise higher throughput, deterministic execution, and tighter integration with the rest of the control system~\cite{barber2023real}. 

\begin{figure}
  \centering
  \includegraphics[scale=0.45]{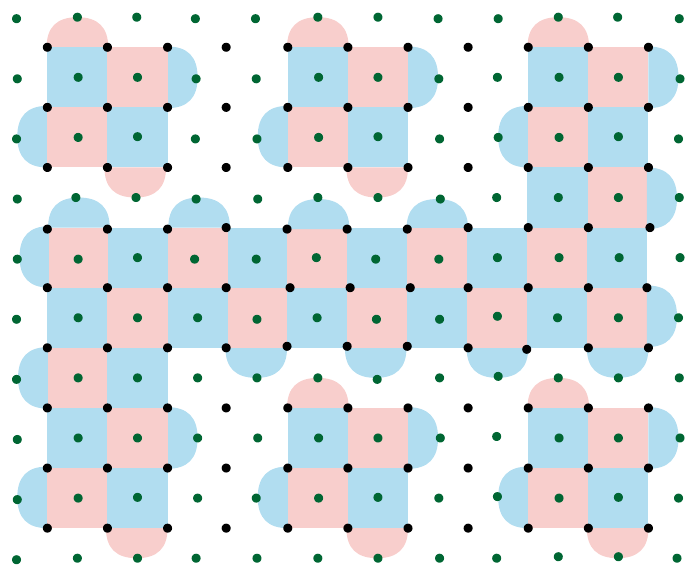}
  \caption{A merge operation in lattice surgery, specifically, a logical $Z\otimes Z$ measurement.}
  \label{fig:lattice_surgery_merge}
\end{figure}

\hlc{Whichever decoder one chooses, a single monolithic instance will inevitably struggle to meet the decoding demands that arise during multi-qubit logic gates.} For surface codes, lattice surgery~\cite{horsman2012surface,fowler2018low,lao2018mapping,litinski2019game,litinski2018lattice,tan2024sat} is the most efficient way to perform multi-qubit logical operations. When lattice surgery is performed on two or more (possibly distant) patches of logical qubits, they are merged into a single patch for multiple measurement cycles before being split apart; a snapshot of the procedure is shown in \Cref{fig:lattice_surgery_merge}. \hlc{It's worth noting that current state of the art decoders can only meet the throughput demands on a patch that is the size of an individual logical qubit up to distance $20$ to $30$}~\cite{barber2023real}. 
\hlc{Worse, their throughput is inversely proportional to the size of the patch}~\cite{higgott2023sparse}. Meanwhile, lattice surgery requires merging multiple such patches~\cite{litinski2019game} and decoding the merged patch.

Dividing the decoding task into overlapping windows is a promising approach~\cite{skoric2023parallel,tan2022scalable,bombin2023modular} to manage this scalability challenge. For instance, prior work~\cite{skoric2023parallel,tan2022scalable} leverages temporal parallelism by processing the syndrome data from many measurement rounds with \textit{temporally} parallel windows. This limits the growth of syndrome backlog when the inner decoder is slower than syndrome generation in terms of throughput. But this strategy doesn't mitigate the large spatial window that each parallel decoder must cover. In ~\cite{skoric2023parallel}, \hlc{the authors introduce the idea of dividing the decoding problem in both time and space, but do not analyze its feasibility.} In~\cite{bombin2023modular}, the authors leverage a different kind of spatial parallelism, where independent subgraphs arising from the logical circuit are decoded in parallel. This addresses the spatial challenge, but requires a dynamic choice in where parity information is routed. We suspect this dynamic routing may prove complicated for hardware decoders to realize in practice.

Here, we explore \textit{spatially parallel windows}. To the best of our knowledge, ours is the first analysis of a decoding scheme that is both capable of handling large patches that arise during logical operations \textit{and} compatible with practical system level constraints of hardware accelerators. As alluded to in ~\cite{skoric2023parallel},  we employ the strategy of dividing the decoding task into multiple overlapping windows, and assign a decoder module to each window. \hlc{The inner decoder that operates on an individual window can be any real time decoder; our scheme is agnostic to the choice. The important point is that these decoders can be coordinated to output a correction when a merged patch spans multiple windows. This technique works because we only need to protect logical information up to the distance of the isolated logical qubits. This allows us to overlap windows with local views and focus on resolving the disagreements at their seams. The decoders are scheduled to run during different time steps, so that they only need to resolve with their neighbors before and after they run.}

A motivating principle of our work is that various design choices need to be carefully balanced to ensure that a scheme with spatially parallel windows can meet the requirements of real time decoding. A hardware constraint specific to the problem is that each window needs to be pre-assigned to an area on the device. This is because real time decoders that operate on individual windows, regardless of the specific implementation, require hardware accelerators with fixed positions. Furthermore, our work emphasizes the importance of avoiding larger windows than necessary, since large windows challenge the scalability of inner decoders. 



We begin by demonstrating how to configure spatially parallel windows in a manner that (1) is compatible with hardware accelerators like FPGAs and ASICs, and (2) can handle the larger merged patches that arise during lattice surgery operations. The decoding scheme is scalable, in the sense that its speed and resource requirements are independent of the number of patches that are merged or how far they are apart on the device. Then we examine the factors influencing the performance, focusing specifically on two key aspects: accuracy and throughput. We proceed to analyze how to achieve a balance between these requirements. Finally, we estimate the size of the individual code patches this scheme can support, assuming one uses an FPGA-based inner decoder.

By performing numerical simulations and analyzing the mechanisms through which spatially parallel windows lead to extra decoding errors, we identify that the size of the windows and the width of the overlapping areas (\hlc{\textit{buffer width}}) are important factors that determine the accuracy. To maintain the fidelity of the logical qubits, the size of the windows cannot be smaller than the individual patches that encode the logical qubits. The buffer width, however, is a key variable in the trade-off between accuracy and throughput. Enlarging the buffer results in larger windows, which subsequently leads to lower throughput (or significantly raises the requirement on computing resources). Conversely, when the buffer is too narrow, it compromises the accuracy of the decoding. We find that the optimal choice of buffer width not only depends on the size of individual code patches, but also depends on the level of physical noise on the device. This can be explained by examining the terms in the logical error rate expression, especially the entropic factors. At a modest physical noise level, the buffer width should be between one-half and two-thirds of the width of an individual patch of code.

The throughput of the decoding scheme is determined by its slowest component. To identify the bottleneck, we separately estimate the speed of both the inter-window communication and the decoding modules. Our findings indicate that the inter-window communication will not become the bottleneck, even when the width of the windows exceed 100. This means the overall throughput of the decoding scheme meets the requirement for real time decoding if and only if each inner decoder module runs sufficiently fast on its window. We discuss the scalability of two state-of-the-art real time decoders~\cite{liyanage2023scalable,barber2023real} when applied to individual windows.

Although we focus on surface code in this work, our results are also relevant to other local codes, e.g. color codes.

%% file: sections/background.tex
\section{Background}
\subsection{Surface code, \hlc{lattice surgery}, and decoding}\label{subsec:bg_surface_code}
\begin{figure}
\centering
\subfloat[]{\includegraphics[scale=0.6]{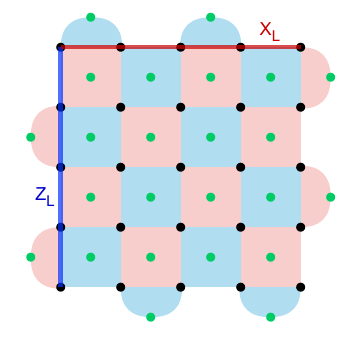}\label{subfig:surface_code}}\qquad
\subfloat[]{\includegraphics[scale=0.6]{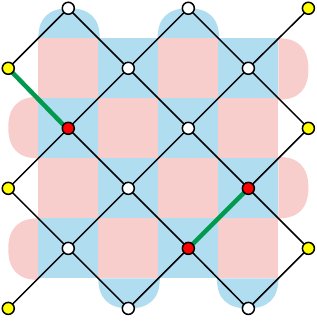}\label{subfig:bg_decoding_graph}}
\caption{(a) A square patch of surface code. The red and blue faces are the X and Z stabilizers, respectively. The blue and red lines mark $\hat{Z_L}$ and $\hat{X_L}$, the $Z$ and $X$ logical operators. (b) An example decoding graph for the X errors. The vertices are the $Z$ stabilizers, augmented by the virtual boundary node (in yellow). The stabilizers that are flipped are marked in red, and a minimum weight matching is denoted by the green edges.}
\label{fig:surface_code}
\end{figure}

A square patch of surface code encodes one logical qubit in a $d\times d$ grid of data qubits, where $d$ is the code distance, and uses $d^2 -1$ syndrome qubits for the stabilizer measurements. A distance-5 example is shown in \Cref{subfig:surface_code}. The stabilizer measurements project errors on the data qubits into Pauli $X,Y$ or $Z$ errors. Each $X$ ($Z$) stabilizer measures the $X$ ($Z$) parity of the data qubits that it acts on, and detects the $Z$ ($X$) errors on them since the $X$ and $Z$ operators anti-commute. An $X$ ($Z$) syndrome qubit reads $1$ if an odd number of its data qubits are affected by a $Z$ ($X$) error. In this case we say the stabilizer is flipped. A flipped syndrome readout is also called an \textit{anyon} or a \textit{defect} in literature. A $Y$ error can be viewed as the product of an $X$ error and a $Z$ error. It can be detected by both types of the stabilizers.

An $X$ ($Z$) logical operator, $\hat{X_L}$ ($\hat{Z_L}$), is a product chain of $X$ ($Z$) operators on a set of data qubits that connect the two $X$ ($Z$) boundaries. Examples of them are shown in \Cref{subfig:surface_code}. The logical operators do not flip any stabilizer. Note that if a chain of operators is equivalent to a linear combination of the stabilizers, it does not flip any stabilizer either, but it acts trivially on the logical qubit. We define the code distance for $X$ ($Z$) errors, $d_X$ ($d_Z$), as the weight of the shortest $X$ ($Z$) logical operator. It is the minimum weight of a nontrivial $X$ ($Z$) error chain that cannot be detected by the code. The code distance $d$ is the minimum of $d_X$ and $d_Z$. If a device has biased noise, e.g. more Z errors than X errors, then a rectangular patch with different $d_X$ and $d_Z$ could be used to further suppress the dominant type of errors.

\hlc{Lattice surgery implements logical multi-qubit Pauli measurements via the merging and splitting of patches, which can be used for performing logical gates. For instance, when two patches are merged along boundaries that coincide with their $Z$ logical operators, the value of $Z \otimes Z$ can be inferred from the product of the Z stabilizers spanning the boundary. $X$ Pauli products are measured similarly, while measurements that involve the Y operator can be done with twist-based lattice surgery}~\cite{litinski2018lattice} \hlc{or an alternative protocol}~\cite{chamberland2022universal}. \hlc{When logical qubits are separated in space, the merge operation can be facilitated by a long ancilla patch as in Figure 5 in}~\cite{litinski2018lattice}.

While all errors chains with weight smaller than $d$ are detectable by the code, only the ones with distance no more than $\lfloor\frac{d}{2} \rfloor$ are guaranteed to be correctable. The decoding of syndrome is performed on a decoding graph (\Cref{subfig:bg_decoding_graph}). For the surface code, independent $X$ and $Z$ decoding graphs can be used for correcting $X$ and $Z$ logical errors. On a decoding graph, each edge represents an error mechanism that is detectable by its vertices. An edge between two adjacent $Z$ stabilizers represents an X flip on the data qubit shared by them. An edge connected to the virtual boundary node represents an error that is only detected by one stabilizer.

For simplicity, \Cref{subfig:bg_decoding_graph} only shows a 2D decoding graph for one round of syndrome measurement. The full decoding graph that a decoder works with has a time dimension because measurement errors also need to be represented. Such a graph can be viewed as multiple copies of \Cref{subfig:bg_decoding_graph} stacked on top of each other. If two vertices represent the readouts of the same stabilizer from adjacent time steps, they are connected by an edge that represents a flip on the syndrome qubit.

A popular technique for decoding syndromes from a surface code is minimum-weight perfect matching (MWPM)~\cite{dennis2002topological}, which identifies the lowest weight error patterns on a given decoding graph. Sparse Blossom~\cite{higgott2023sparse} and Fusion Blossom~\cite{wu2023fusion} are recent, fast software implementations of this technique. Union-Find (UF)~\cite{delfosse2020linear} is another popular technique, and can be viewed as an approximation of the Blossom algorithm~\cite{edmonds1965maximum,edmonds1965paths} that implements MWPM~\cite{wu2022interpretation}. It is known to have better runtime complexity and slightly lower accuracy than MWPM.
\subsection{Real time decoding with FPGA}\label{subsec:bg_fpga}
For concreteness, we explore the implications of using an FPGA for the inner decoder. FPGAs offer an attractive alternative to general-purpose CPUs for tasks with strict timing requirements and modest programmability requirements.
An FPGA is made up of a ``fabric'' consisting of a great many components whose behaviors are individually programmable, whose input and output can be flexibly routed among other components, and whose timing is uniformly specified so as to support easy latching of inter-component signals.
These properties imbue FPGAs with excellent parallelism and pipelinability, enabling them to achieve excellent throughput on high-bandwidth tasks.
The primary trade-off for these abilities is that the FPGA fabric is almost always programmed ``once and for all''---that is, the individually programmable components nonetheless have their behaviors fixed for the lifetime of an application.

Because surface code syndromes arrive in a stream at approximately $(d^2 - 1)$ Mbps, decoding is a strong candidate application for an FPGA~\cite{liyanage2023scalable,das2022lilliput,vittal2023astrea}.
However, the constraints of FPGA programming bear directly on the decoding problem:
\begin{itemize}
    \item Scalability:
    An individual FPGA enjoys a fixed set of components, making it suitable for decoding only up to a certain size.
    Meanwhile, arbitrarily long swaths of surface code may appear during lattice surgery protocols.
    \item Communication:
    The high-speed communication afforded by the lock-step evolution of FPGA components is somewhat lost when connecting two separate fabrics together (e.g., over ethernet).
    Accordingly, programmers need to keep tight control over inter-fabric communication to keep it from dominating the runtime of an application.
    Additionally, local connections between components are set out at ``compile time'', sometimes coming at significant layout cost.
    \item Fixed gateware:
    Since FPGAs only permit modest on-the-fly reprogramming, care must be taken to match these limited abilities with the changing decoder requirements of a surface code patch undergoing lattice surgery.
\end{itemize}
Deploying an FPGA to accomplish decoding requires accommodating each of these constraints.

%% file: sections/related_work.tex
\section{Related work}\label{sec:related_work}
In recent years, there has been a surge in research on real time decoding for QEC. Helios~\cite{liyanage2023scalable} is an FPGA-based UF decoder. When acting on a surface code with distance $d$, it achieves a sublinear average time complexity per cycle, at the cost of $O(d^3)$ hardware resources. The work demonstrated an implementation operable up to $d = 21$. Riverlane~\cite{barber2023real} recently implemented a UF decoder on both FPGAs and ASICs and reported results for up to $d=23$. Astrea ~\cite{vittal2023astrea} and LILLIPUT~\cite{das2022lilliput} output the same solution as a MWPM decoder for surface code up to $d=7$ and $d=5$, respectively. These can be implemented on FGPAs. AFS proposes to implement the UF decoder on ASICs~\cite{das2022afs}. There are also real time decoders with SFQ-based superconducting digital circuits~\cite{ravi2023better,holmes2020nisq+,ueno2021qecool}. Some decoders use hierarchical decoding~\cite{ravi2023better,chamberland2023techniques,smith2023local}.

A few recent works divide the decoding task into overlapping windows~\cite{skoric2023parallel,tan2022scalable,bombin2023modular}, an approach that is also employed in our work. In~\cite{tan2022scalable} and~\cite{ skoric2023parallel}, the authors \hlc{primarily address parallelization in time, which helps contain the backlog when the throughput of the decoder does not match the rate that the syndrome is generated.} In~\cite{skoric2023parallel}, the authors \hlc{also includes a discussion on dividing the decoding problem in space but do not explore the consequences.} The information passed between neighboring windows is similar, whether they are temporally or spatially parallel. However, temporally and spatially parallel windows apply to different problems and so have different constraints and goals. For example,~\cite{skoric2023parallel} does not investigate how a narrow buffer would compromise the accuracy of decoding.

A paper from PsiQuantum~\cite{bombin2023modular} is the only prior work that targets lattice-surgery style fault-tolerant blocks. Its focus is different from our work though. It prioritizes minimizing the latency of a software implementation and does not address ``further hardware and systems considerations that are relevant". However, we put more emphasis on the constraints of hardware accelerators, and consider throughput instead of latency as the more important measure of speed for real time decoders. For example, they develop algorithms that take lattice surgery blocks as input, then assign windows, while we explicitly require that windows have pre-assigned positions so that they are compatible with hardware accelerators. Both papers include numerical results on how the buffer width affects decoding accuracy, but we provide more analysis and also study how the influence depends on the physical noise level.



%% file: sections/parallel_windows.tex
\section{framework of the decoding scheme}\label{sec:framework}
In \Cref{subsec:bg_fpga}, we discussed three constraints of an FPGA implementation. Our protocol keeps these constraints foremost in mind: our starting premise is that a single decoder handles a fixed (small) window; we explicitly analyze the single burst of communication needed from one stage of decoding to the next, which takes place over a fixed subset of a nearest-neighbor topology network; and we limit our reprogrammability requirements to a simple form of ``masking'', where read and write operations are individually dis/allowed at given cells.

In this section we explain key aspects of the design. We first show how the decoder modules should be connected and why it is necessary for them to overlap in buffer regions. Then we discuss the metrics of the decoding scheme and the design choices that influence them. Finally we introduce a configuration that is compatible with the hardware constraints.

\subsection{Connecting decoder modules}\label{subsec:framework}
The time that it takes for a monolithic decoder to process a long patch of code scales at least linearly with the area of the patch. \hlc{(Exceptions like look-up-table-based decoders have exponentially high storage overhead so they only work for very small patches}~\cite{das2022lilliput}). Spatially parallel decoding circumvents this by dividing the decoding task into multiple processes. In order to avoid frequent communication between the processes, we do not consider fine-grained schemes where each (ancilla or data) qubit is assigned its own decoder module. We only consider coarse-grained parallel schemes where each decoder module works on an area that is similar to an individual patch of logical qubit like \Cref{fig:surface_code}(a). We refer to the area that a decoder module acts on as a \textit{window}.

\begin{figure}
  \centering
  \subfloat[]{\includegraphics[scale=0.5]{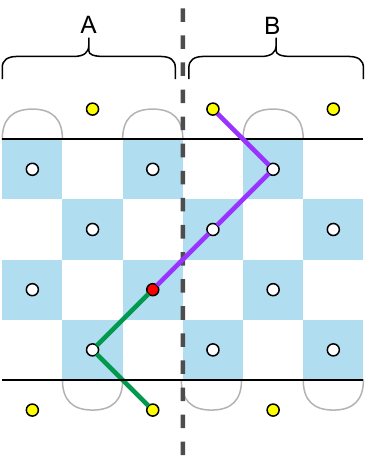}
  \label{subfig:no_buffer1}}
  \hspace{1cm}
  \subfloat[]{\includegraphics[scale=0.5]{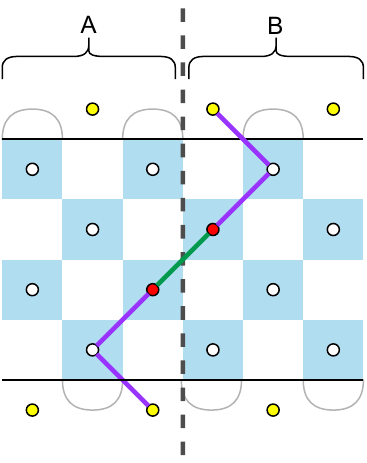}
  \label{subfig:no_buffer2}}
  \caption{\hlc{A segment from a longer patch of surface code, divided into 2 non-overlapping windows by an artificial boundary (dashed line). The optimal corrections are shown in green and the suboptimal ones in purple. (a) One flipped syndrome in window A. The suboptimal correction will be selected if matching to the artificial boundary is allowed.  (b) An error flips two syndromes, one in each window. The suboptimal correction will be selected if matching to the artificial boundary is disallowed.}}
  \label{fig:no_buffer}
\end{figure}

It might seem resource efficient to divide the device into non-overlapping windows, and run decoders on each window simultaneously. However, this will either completely sacrifice the accuracy of decoding, or require high communication overhead between neighboring decoder modules. \hlc{If no communication is allowed between non-overlapping windows, the decoder has two choices when acting on one window: (1) it allows matching across the boundary with neighboring windows, (2) it does not allow such matching. The examples in} \Cref{fig:no_buffer} \hlc{show that either case leads to easy decoding failures. In case (1), the flipped syndrome in Figure 3(a) would be matched to the boundary between the two windows, while it should be matched to the lower boundary. In case (2), the flipped syndromes in Figure 3(b) would be matched to the top and bottom boundaries, while they should be matched to each other. Both examples result in logical errors.}

In order to maintain the accuracy of decoding and avoid high communication overhead, one can take the approach of using overlapping windows and applying decoder modules on neighboring windows in different \hlc{\textit{layers} (time steps). We will use terms \textit{rough boundary} and \textit{smooth boundary} in the explanation. On a decoding graph, a boundary with edges to the virtual boundary node is termed rough, while a boundary with no edges out of it termed smooth. These boudaries can refer to either ones that already exist in the global decoding graph (e.g. the top and bottom boundaries of Fig. 4 are rough), or an artificial one that arises from partitioning the patch into multiple windows (e.g. the vertical boundaries in Fig. 4 are rough and smooth respectively). These terms should not be confused with the rough and smooth boundaries of a surface code, which are synonymous with boundaries that absorb strings of Z and X flips, respectively.}

\begin{figure}
\centering
\sidesubfloat[]{%
  \includegraphics[clip,scale=0.5]{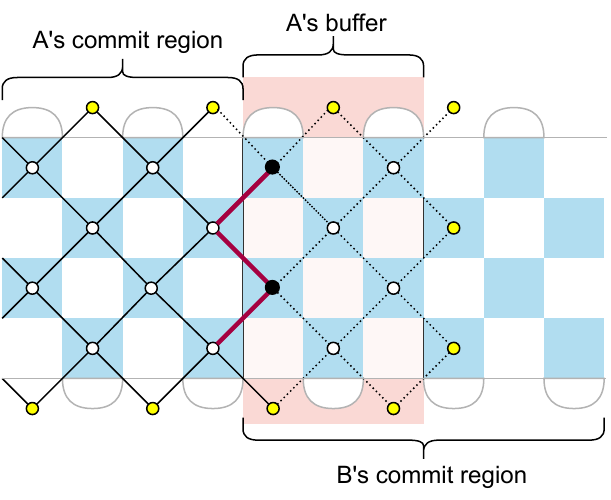}%
  \label{subfig:rough_boundary}
}\\
\sidesubfloat[]{%
  \includegraphics[clip,scale=0.5]{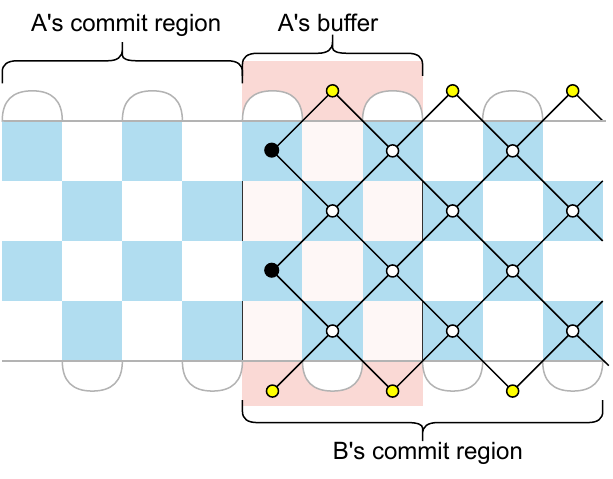}%
  \label{subfig:smooth_boundary}
}
\caption{Decoding graphs for $X$ errors, used by (a) window A and (b) window B. In both (a) and (b), the buffer is marked in pink, and the two nodes where artificial defects could be introduced are marked by solid black dots. In (a), the red edges are the ones that connect the nodes in A's commit region to the buffer. Corrections along the dotted edges are on nodes in the buffer, and are not committed by window A. In (b) the decoding graph has solid edges in the buffer, because the corrections in the buffer are finalized in this step.}
\label{fig:decoding_graphs_pw}
\end{figure}

\Cref{fig:decoding_graphs_pw} illustrates this approach \hlc{for a segment of the X decoding graph taken from a longer patch. As shown in shown in Figure 4a, window A has a rough boundary on its right side. The edges that connect to this boundary are derived from those that straddle window A in the global decoding graph. In the first layer, only window A is acted on by a decoder. After it obtains a correction, the decoder only commits the part that acts on the nodes in A's \textit{commit region}, marked by the solid edges in Figure 4a.} The edges with dotted lines are in the buffer between the windows A and B, and corrections in this region are not applied by window A.

The only communication between windows A and B happens when A finishes decoding. And the only information A passes to B is \hlc{the \textit{artificial defects}, the nodes along the left (smooth) boundary of B that are flipped by corrections in the commit region of A. The artificial defects are marked with solid black dots in Figure 4. After B learns about these defects, it performs corresponding flips on the nodes along its left boundary and begins decoding, as shown in Figure 4b. This ensures consistent corrections across the windows.}

\hlc{Unlike in window A, the decoding graph of window B (Figure 4b) does not have an artificial rough boundary with neighboring windows. This is true for any window scheduled in the final layer. In the X decoding graphs shown in Figure 4, there are still natural rough boundaries at the top and bottom; however, these would be smooth in the corresponding Z decoding graph and so this graph would have no rough boundaries. When a decoding graph has no rough boundaries, the decoder will fail to output a correction if the syndrome it receives contains an odd number of flipped nodes.} In other words, there is a \textit{lone defect} that cannot be matched. When this happens, we terminate decoding and report that a logical error has occurred. This failure is due to suboptimal decoding in previous layers.

\subsection{Metrics}\label{subsec:metrics}
Throughput is one of the most important metrics for a real time decoder. It should match the rate at which syndromes are generated \hlc[pink]{so that the syndrome backlog will not grow exponentially, exhausting storage space and slowing down logical operations}~\cite{terhal2015quantum}. When the decoder has enough throughput, it is sufficient to apply the sliding window technique~\cite{iyengar2011windowed} along the time dimension. \hlc[pink]{When the inner decoder is not fast enough, a temporally parallel decoding scheme can mitigate the backlog problem. But temporal parallelization alone is not an ideal solution for the long patches that arise from lattice surgery, which will require more layers of temporally parallel windows. This will slow down the logical clock rate further and also incur more hardware costs if the decoder is implemented with hardware accelerators.}

Accuracy is a critical metric for any decoding scheme. When multiple logical qubits undergo a logical operation, they should ideally maintain the same level of fidelity as when they are isolated. As we will show in \Cref{sec:ler}, this requires sufficiently large windows and buffers.

Latency is another metric of the decoder's efficiency. It measures the total delay between the time that a syndrome measurement cycle finishes and the decoder outputs a correction. \hlc[pink]{It is particularly important when performing conditional operations, like gate-by-measurement. The latency of the decoder determines how long a logical qubit needs to idle before the conditional operation can be executed. This is because the value that conditions the operation is in part determined by the decoder's corrections. While it doesn't have a strict limit like throughput, a large latency is still undesirable because it leads to an increase in the space-time volume of the computation}~\cite{chamberland2023techniques,delfosse2023choose}. \hlc[pink]{Specifically, the time of the computation increases linearly with the latency, and to protect the logical information, the code distance should be increased logarithmically to compensate. 
However, the scheme we consider only has a latency of a few cycles. The logical error rate per cycle is roughly proportional to $(100p)^{(d+1)/2}$}~\cite{fowler2013surface}. \hlc[pink]{At $p\sim 0.1\%$, increasing $d$ by 2 would suppress the logical error rate by $\sim10$X, which is enough for overcoming the latency.}

The latency of the spatially parallel windows is determined by (1) the latency in each component of the decoding scheme, which is already reflected in the throughput, and (2) the number of layers that the windows are divided into. To reduce the latency, it is preferable to use a window configuration with a small number of layers. This is one of the considerations in the following subsection.

\subsection{Window configuration for hardware decoding}\label{subsec:staggered_squares}
For a software implementation of the spatially parallel windows, one can dynamically and flexibly adjust the configuration of the windows during the computation, given the lattice surgery operations at each time step. However, for real time decoding using a hardware accelerator like an FPGA or ASIC, the positions of the windows and the links between them should ideally be fixed. Here, we seek a window configuration that is compatible with hardware accelerators and also accommodates general lattice surgery operations.

Besides having fixed window positions and links, the window configuration and mapping/compiling strategy should respect three other constraints to ensure the accuracy of decoding. First, the commit regions of the windows in the same layer must be disjoint. Unlike windows in different layers (e.g. the two windows in \Cref{fig:decoding_graphs_pw}), they cannot share data qubits. This is because no information is shared between windows in the same layer, while the information necessary for making consistent corrections is passed between neighboring windows in different layers. Second, the windows should not have smaller size than an isolated patch of code (e.g. \Cref{subfig:surface_code}). Thirdly, \hlc{when a patch spans multiple commit regions, the intersection of the patch with each commit region should not be too small.} The second and third points will be further explained in \Cref{sec:ler}.

\begin{figure}
  \centering
  \subfloat[]{\includegraphics[scale=0.35]{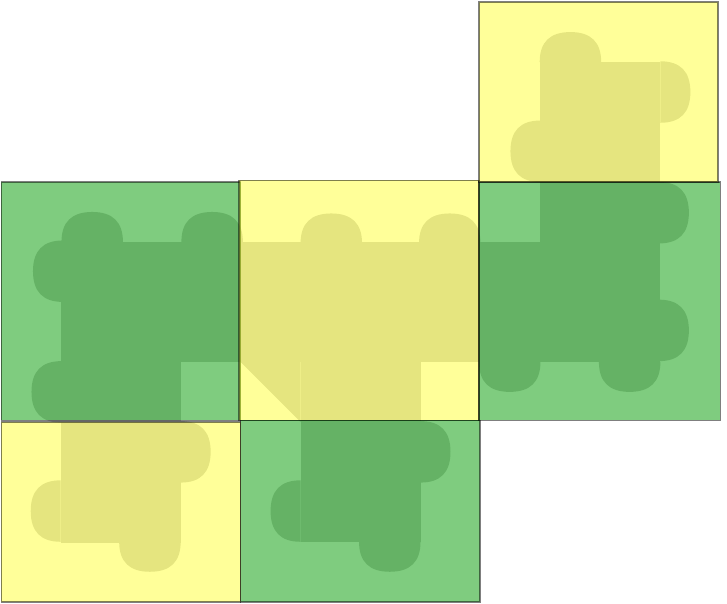}
  \label{subfig:tree}}
  \subfloat[]{\includegraphics[scale=0.35]{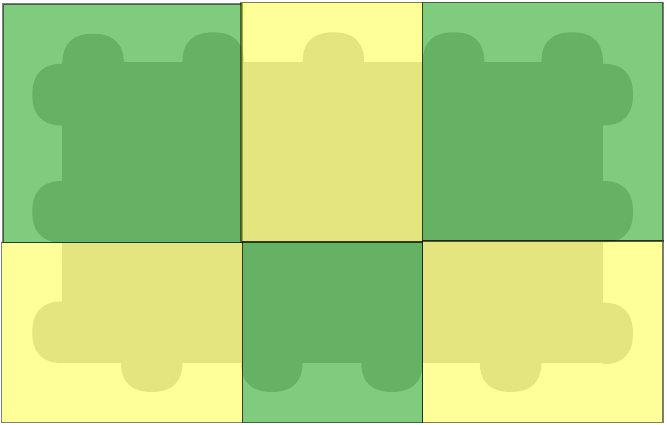}
  \label{subfig:y_meas}}
  \caption{(a) A patch with a tree structure is 2-colorable. (b) A patch that arises during a $Y\otimes Y$ logical measurement. Not 2-colorable.}
  \label{fig:checkerboard}
\end{figure}

To reduce the latency of the decoder, the windows should be arranged in a small number of layers. Can a two-layer configuration meet the requirements? For the large patch in \Cref{fig:lattice_surgery_merge}, it suffices to use a 2-colorable checkerboard configuration of windows. This is because the patch has a tree structure. See \Cref{subfig:tree} for a window configuration that accommodates a multi-patch merge of this type (only the commit region of each window is shown). In this example, although the commit regions with the same color overlap at corners, their committed corrections do not conflict because no active qubits are at these corners. 

However, not all patches from lattice surgery operations have a tree structure similar to that in \Cref{fig:lattice_surgery_merge} and \Cref{subfig:tree}. These examples only involve $X$ and $Z$ logical measurements. But when a lattice surgery operation involves a Y measurement on a logical qubit, the patch may need to be touched on more than one $X$ or $Z$ edge, which may result in a patch that cannot be accommodated by a checkerboard pattern. For instance, \Cref{subfig:y_meas} shows a merged patch that arises during a $Y\otimes Y$ logical measurement~\cite{chamberland2022circuit}.

\begin{figure*}
  \centering
  \subfloat[]{\raisebox{0.38cm}{\includegraphics[scale=0.5]{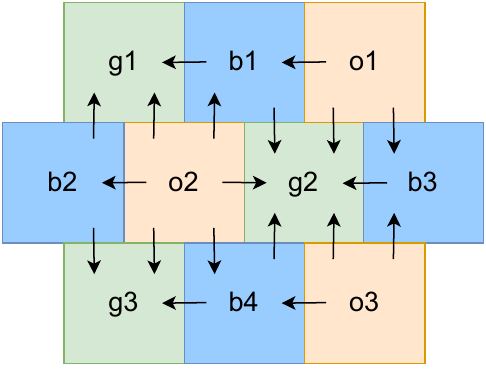}}\label{subfig:staggered_squares_arrows}}
  \subfloat[]{\includegraphics[scale=0.5]{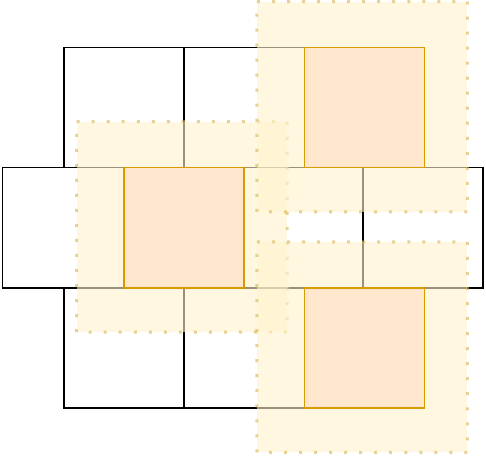}\label{subfig:staggered_squares1}}
  \subfloat[]{\includegraphics[scale=0.5]{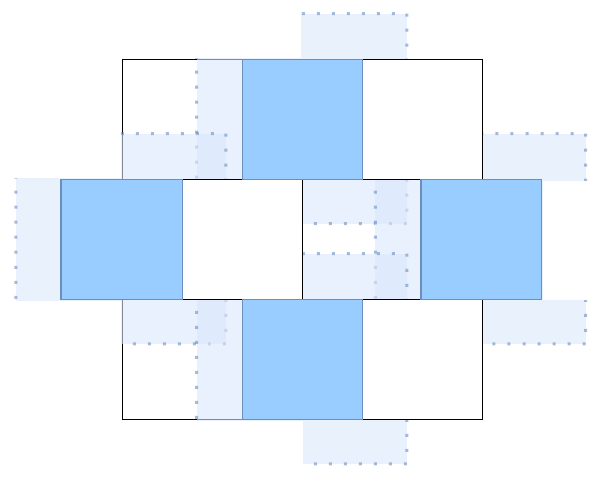}\label{subfig:staggered_squares2}}
  \subfloat[]{\raisebox{0.37cm}{\includegraphics[scale=0.5]{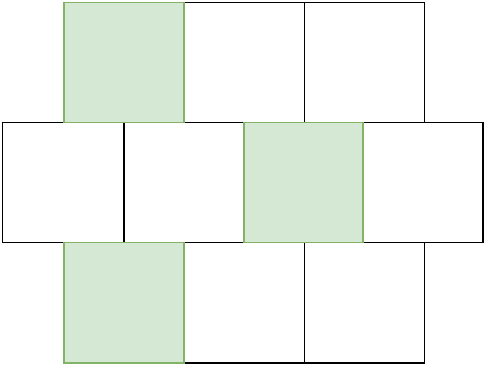}}\label{subfig:staggered_squares3}}
  \caption{A window configuration with staggered squares. (a) Only showing the commit region of each window. Each arrow represents a one-way link that communicates information on artificial defects. (b-d) The windows in the first, second, and third layers, respectively. The buffers are shown in lighter color and surrounded by dotted edges. The windows on the last layer do not have buffers. The buffer width in this figure is chosen to improve readability.}
  \label{fig:staggered_squares}
\end{figure*}

To accommodate general lattice surgery operations we propose to use a 3-coloring of staggered squares (\Cref{fig:staggered_squares}). It is a simple configuration that satisfies all the constraints mentioned above. The size of each commit region is the same as the size of an individual patch of code. If the rows of windows are not staggered then 4 colors are required, which leads to an unnecessary increase of latency.

%% file: sections/logical_err.tex
\section{Logical error rates}\label{sec:ler}
In this section we study the logical error rates (LER) when using spatially parallel windows. We begin by describing the simulation setup on a rectangular surface code patch, and then study logical errors along the short and long edges of the patch. We \hlc[pink]{then move on to simulations and analysis for a more general setup, five windows stacked in two rows.} We also discuss how the results influence design decisions (e.g. buffer width) and introduce constraints on mapping.
\subsection{Methodology for simulations}\label{subsec:methodology}
We perform quantum memory simulations, where each shot includes $d$ cycles of stabilizer measurements. We use Stim~\cite{gidney2021stim} for simulating stabilizer circuits along with a circuit-level noise model that includes single-qubit, two-qubit, and measurement errors characterized by a single parameter $p$. We use PyMatching2~\cite{higgott2023sparse}, a state-of-the-art implementation of the MWPM decoder, both as the inner decoder applied to individual windows and as the baseline global decoder. Although evaluation of the throughput would require a real time inner decoder, PyMatching2 is sufficient for simulations that focus on logical error rates and syndrome densities.

\begin{figure}
  \centering
  \includegraphics[scale=0.4]{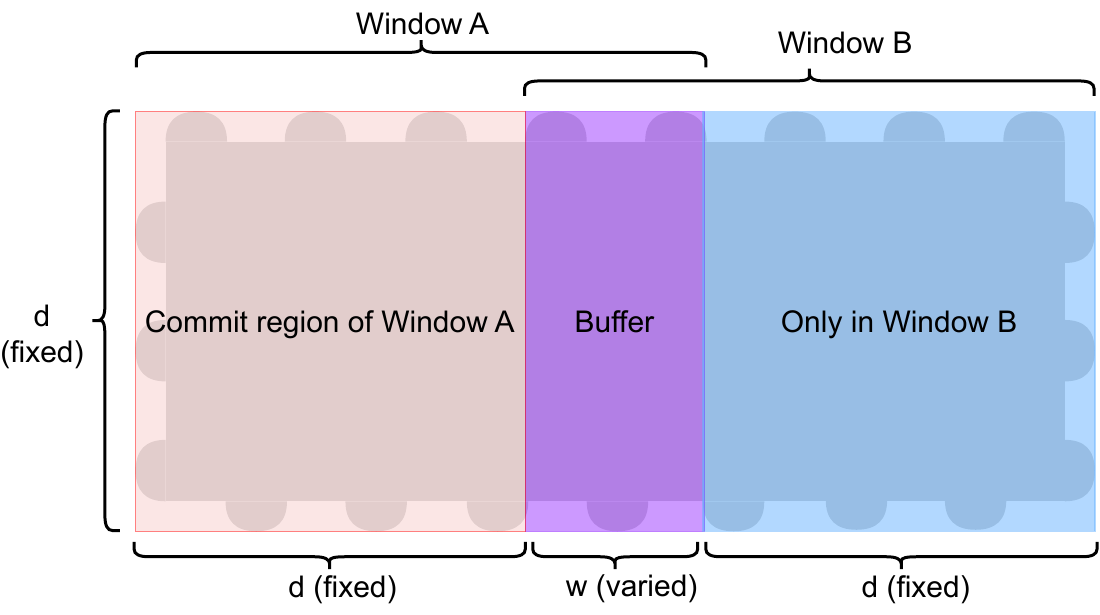}
  \caption{Setup for the numerical simulations.}
  \label{fig:sim_setup}
\end{figure}

For \Cref{subsec:ler_short_edge,subsec:ler_long_edge} we use a rectangular surface code patch that would arise when merging two square-shaped patches of distance $d$ during lattice surgery. Our goal is to maintain the fidelity of the individual patches, and we study the impact of the buffer width and physical noise $p$ on this fidelity. As such, we use the simplest setup (\Cref{fig:sim_setup}) with just two overlapping windows, with window A applied before B. This configuration is such that the square-patches are the same size as the commit regions of the windows. 

\subsection{Logical errors along the short edge}\label{subsec:ler_short_edge}
\hlc{In a memory experiment, a logical error occurs if an X or Z logical operator is flipped after applying both the noise and the correction.} We first study the impact of logical errors along the short edge of \Cref{fig:sim_setup}, with code distance $d = 15$. We only use one of the X and Z (global) decoding graphs: the one \hlc{with rough boundaries on the top and bottom edges. For this setup the logical operator should be a horizontal string that connects the left and right edges of the entire patch.}

\begin{figure}
  \centering
  \includegraphics[scale=0.8]{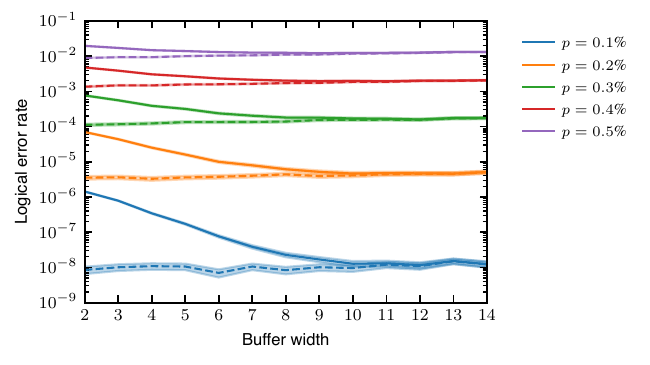}
  \caption{Logical error rate v.s. buffer width, using the setup in \Cref{fig:sim_setup} with $d=15$, counting the logical errors along the short edge (that connect the top and bottom boundaries). The dashed lines are the baseline results from using a \textit{global} decoder on the same underlying patch. The shaded regions indicate the $95\%$ confidence intervals. Each data point at $p=0.1\%$ is obtained with 8B shots.}
  \label{fig:short_edge_ler}
\end{figure}

In \Cref{fig:short_edge_ler} we plot the LER against $w$, \hlc{the buffer width (e.g. Figure 4 has $w=3$)}. The figure shows that when the buffer width is small, the accuracy of the parallel decoder is orders of magnitude worse than its global counterpart. It also shows that increasing $w$ closes the gap. In fact, the gap diminishes before the buffer is grown to match the length of the short edge, which implies that $w$ need not be as large as $d$ to maintain accuracy. This is beneficial for throughput, since large buffers invariably decrease the speed of each decoder module or place tighter requirements on the underlying hardware. 

So how large does $w$ need to be? From \Cref{fig:short_edge_ler} we observe that the optimal choice of $w$ depends on the physical noise $p$. At $p=0.1\%$, the $95\%$ confidence intervals of the LERs from parallel and global decoding overlap when $w\geq 10$. At $w=10$ and $p=0.1\%$, the ratio of parallel to global LER is $\sim 1.32$. But at $p=0.5\%$, the ratio is below $1.3$ when $w\geq 6$, and it is only $\sim 1.05$ at $w=10$. This shows that one might need a larger $w$ at lower $p$ to match the accuracy of global decoding.

\begin{figure}
\raggedright
\sidesubfloat[]{
  \includegraphics[clip,scale=0.76]{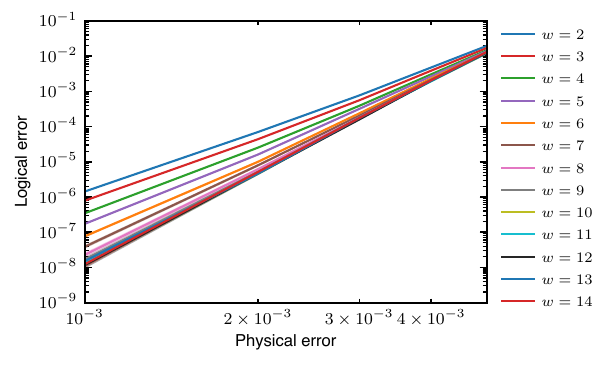}
  \label{subfig:pw_slopes}
}\\
\sidesubfloat[]{
  \includegraphics[clip,scale=0.8]{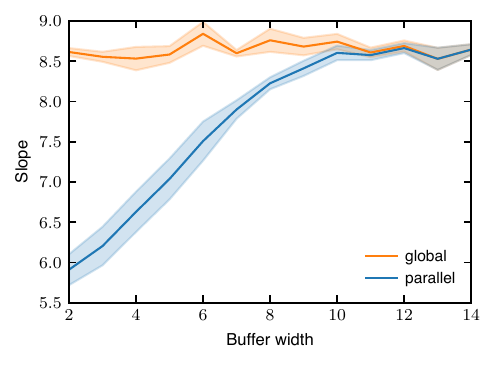}
  \label{subfig:slope_v_w}
}
\caption{The data in \Cref{fig:short_edge_ler} represented differently. (a) Logical error rate v.s. physical error rate, for parallel decoding with different buffer width $w$. 
(b) The slopes of the lines in (a) and (b), plotted against $w$.}
\label{fig:short_edge_slopes}
\end{figure}

In \Cref{fig:short_edge_slopes}, we provide an another view of the results from \Cref{fig:short_edge_ler}. In \Cref{subfig:pw_slopes} we plot the LERs against $p$ on log scale. The slope of a line in the plot shows how quickly the LER is suppressed when $p$ decreases, and is a proxy for effective code distance. When we plot the slope against $w$ in \Cref{subfig:slope_v_w}, we see that the benefit of growing $w$ diminishes as $w$ increases.

\begin{figure}
  \centering
  \subfloat[]{\includegraphics[scale=0.35]{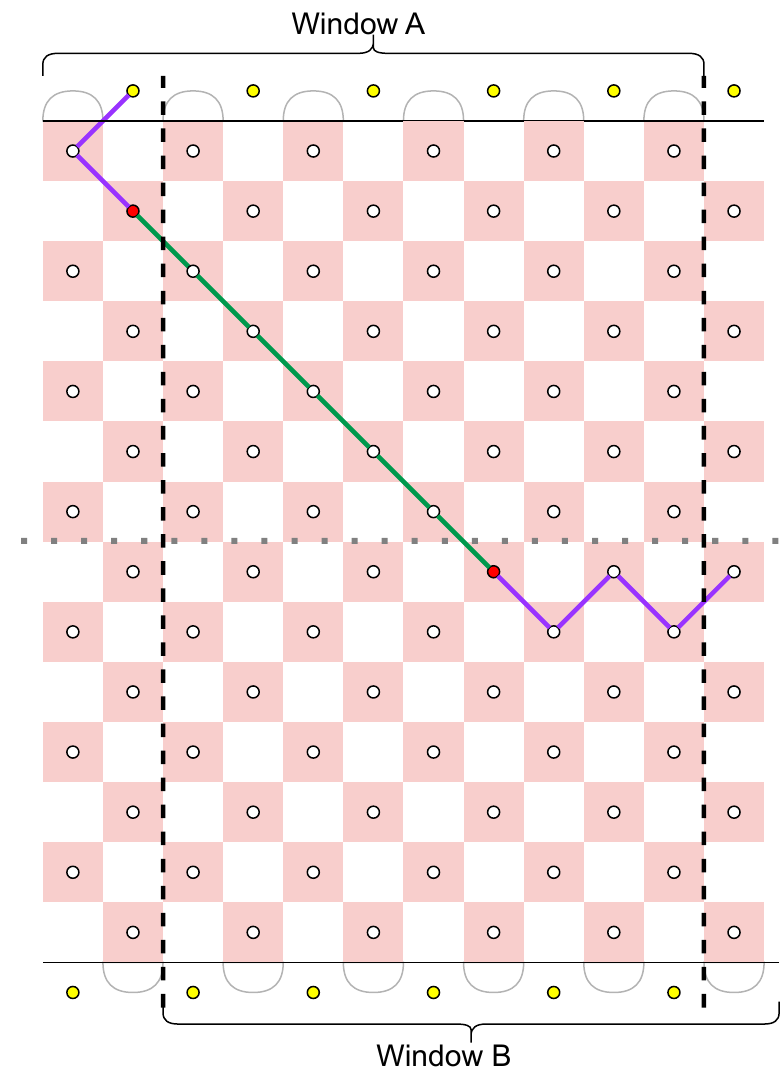}
  \label{subfig:err_str_w9}}
  \hspace{3mm}
  \subfloat[]{\includegraphics[scale=0.35]{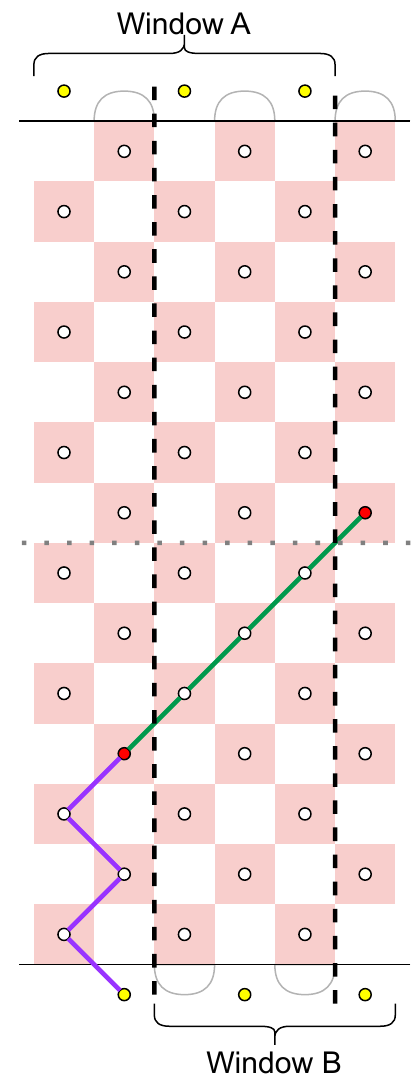}
  \label{subfig:err_str_w3}}
  \caption{Example error strings that confuse the parallel decoder but can be corrected by a global decoder. Left: $w=9$. Right: $w=3$. In each example, the region between the two dashed vertical lines is the overlap of windows A and B. Window A is decoded first. The flipped stabilizers are in red. The error strings are in green. The purple edges are wrong decoder outputs by A. The dotted horizontal line marks the middle of the patch.}
  \label{fig:err_str_short_edge}
\end{figure}
 
To facilitate understanding of the numerical results, we also study error strings that confuse the parallel decoder but can be corrected by a global decoder. In \Cref{fig:err_str_short_edge} we show examples of such error strings for $d=15$, $w=9$ and $3$. The example error strings have lengths ($l$) 6 and 4, respectively. Since $l$ is no more than  $\lfloor \frac{d}{2}\rfloor$ (which is 7) for both strings, they can be corrected by a global decoder. 

However, in \Cref{subfig:err_str_w9}, window A can match the left end of the string to the top boundary (and commit this), and the right end to the rough boundary outside of the buffer (tentatively). Such a matching consists of $6$ edges, the same as the correct one, so window A will output one or the other with equal probability. When window B is in action, it only sees the right end of the error string, and matches it to the bottom boundary. Since one end of the error string is matched to the top boundary and the other is matched to the bottom, \hlc{the correction flips the horizontal logical operator while the noise does not. Thus the operator is flipped at the end and} we have a logical error. In \Cref{subfig:err_str_w3}, window A can match the left end of the error string to the bottom boundary. The right end of the string is invisible to A, so A does not need to act on it. Such a matching is again the same weight as the actual error string. Then window B only sees the right end of the error string and matches it to the top boundary.

How long do these strings need to be? For a connected error string (that is correctable by some global decoder) to result in a logical error along the short edge, it should satisfy all the three conditions below:
\begin{enumerate}
    \item One end of it is matched to the top boundary and the other is matched to the bottom, in the committed matching
    \item Exactly one end of it is in the commit region of window A. If neither end is in the commit region of A, no correction will be committed by window A, and the error is entirely decoded by B as in the case of a global decoder. If both ends are in the commit region of A, then the corrections to both defects are committed by A.
    \item In window A, the end that is not matched to the top/bottom boundary is matched to the rough boundary out of the buffer. If both ends are matched to the same boundary, we do not get a logical error. If they are matched to opposite boundaries by A, the global decoder would also do such a matching.
\end{enumerate}

Let $l$ be the weight of the error string (its number of edges). The right end of the error string can be just above or just below the middle of the patch, and the left end can be just within the commit region of A. The weight of the wrong matching in A, described above, is the sum of the distance from the right end to the rough boundary out of the buffer, and from the left end to the top/bottom boundary. Assuming an odd $d$, the first part is $\max (w+1-l,0)$, and the second part is $\max (\frac{d-1}{2}+1 - l, 0)$, for a diagonal error string similar to the ones in \Cref{fig:err_str_short_edge}. The wrong matching might be output by decoder A if it has no higher weight than any correct one. Then for $d=15$, a string with $l=4$ ($5,6,7$) can confuse a parallel decoder with $w\leq 3$ ($w\leq 6, w\leq 9, w\leq 12$). Note that when $w \geq d-2$, a connected error string that can confuse the parallel decoder is also long enough to confuse any global decoder.

The lengths and the quantity of error strings help explain the LER. The chance of having a weight $l$ error is roughly $(1-p)^{n-l} p^{l}$, where $n$ is the total number of edges in the 3D decoding graph (we say \textit{roughly} because the probability associated with an edge in the decoding graph is not exactly $p$ --- the circuit-level noise model adds complications). The LER can be expressed as $P = \sum_{l=l_{\min}}^n N_{\text{fail}}(l)(1-p)^{n-l} p^{l}$, where $N_{\text{fail}}(l)$ is the number of weight-$l$ errors that cause logical error and $l_{\min}$ is the minimum weight of an uncorrectable error (which does not need to be a connected string). 

When $p$ is low, the LER is usually dominated by the term from $l_{\min}$. This is why $l_{\min}$ is an important quantity, whether in the decoder-specific context in this paper or the decoder-independent context where it is equivalent to $\lceil \frac{d}{2} \rceil$. The range of $p$ we use for the numerical simulations includes $0.1\%$ which is already low. However, our simulation results cannot be fully explained by the min weight of uncorrectable errors. For example, \Cref{subfig:err_str_w3} has a $l_{\min}$ of 4 with parallel decoding, but it has more than two orders of magnitude lower LER than a distance 8 surface code decoded globally (which also has $l_{\min}=4$) under the same $p$ of $0.1\%$.

The value $l_{\min}$ alone does not fully explain the simulation results because the entropy of error chains sometimes play a more important role in determining the LER at modest physical noise~\cite{beverland2019role,bravyi2013simulation}. When the buffer is just narrow enough for an error of weight $l_{\min}$ to cause a logical error, as in both examples in \Cref{fig:err_str_short_edge}, \hlc{the entropic factor} $N_{\text{fail}}(l_{\min})$ is a tiny value. The 
entropic factors of higher order terms in $P$ are significantly larger. 
When $p$ is not low enough, the disparity in error probability is not enough to offset the difference in the entropic factors, then the main contribution to the LER is from the higher order terms. If the LER from the parallel decoder is dominated by terms with weight no less than $\lfloor \frac{d}{2} \rfloor$, then it will be close to the LER from the global decoder, even when the two decoding schemes have different $l_{\min}$. This explains the results in \Cref{fig:short_edge_ler}.

\subsection{Logical errors along the long edge}\label{subsec:ler_long_edge}
For a rectangular patch of surface code with a long edge and a short edge, the logical errors along the long edge constitute only a small portion of the total LER when a global decoder is used. Here, we check that spatially parallel windows preserve this property.

\begin{figure}
  \centering
  \includegraphics[scale=0.8]{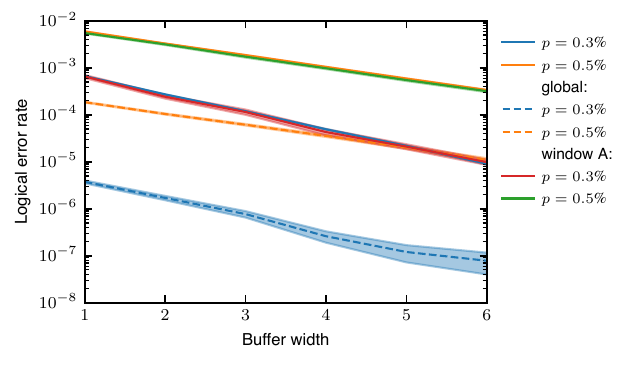}
  \caption{Logical error rate v.s. buffer width, using the setup in \Cref{fig:sim_setup} with $d=7$, counting the logical errors along the long edge. The dashed lines are the baseline results from using a \textit{global} decoder on the same patch. \hlc{The results from applying the global decoder on a patch that is the size of window A are also shown.}}
  \label{fig:long_edge_ler}
\end{figure}

We again use the setup in \Cref{fig:sim_setup}, and show the results of numerical simulations in \Cref{fig:long_edge_ler}. This time, we observe that the LER gap between parallel and global decoding stays constant as the buffer is grown. We note a distinction between \Cref{fig:long_edge_ler} and \Cref{fig:short_edge_ler}. As the buffer width $w$ increases, the code distance increases along the long edge but remains constant along the short edge. This is why the LER from the global baseline decreases as $w$ grows in \Cref{fig:long_edge_ler} but remains almost constant in \Cref{fig:short_edge_ler}. We also find that the LER from parallel decoding coincides with the one from applying the global decoder to a smaller patch that is the same size as window A.

\begin{figure}
  \centering
  \includegraphics[scale=0.48]{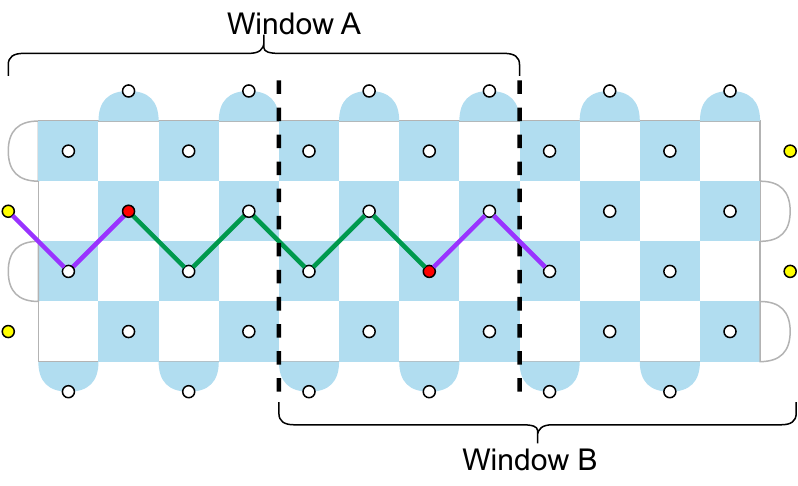}
  \caption{An example error string that causes a logical error along the long edge, when the parallel decoder is used. As in \Cref{fig:err_str_short_edge}, window A is applied before B, the buffer is between the vertical dashed lines, the error string is in green, and a wrong decoder output by A is in purple.}
  \label{fig:err_str_long}
\end{figure}

Along the long edge, we must consider error strings like the one in \Cref{fig:err_str_long}. Window A in this figure has code distance 9, the entire patch has distance 13, and the green error string has weight 5. The string is not long enough to confuse the global decoder on the entire patch, but is long enough to confuse a window. In particular, window A would prefer the incorrect purple matching to the green one, because it only has weight $4$. Then window B, which only sees the right end of the error string, has no choice but to match it to right rough boundary of its decoding graph. The result is a logical error along the horizontal edge.

Our findings about the LER along the long edge are conclusive: 
as long as the horizontal dimension of window A in \Cref{fig:sim_setup} is longer than $d$ (which will be the case due to the buffer), the increased LER from logical errors along the longer edge is negligible when compared to the increase from the errors along shorter edge. Therefore we do not need to consider this type of logical errors when choosing $w$.

\subsection{Generalizing to 2D configurations}\label{subsec:sim_2d}
\hlc[pink]{As mentioned in }\Cref{subsec:staggered_squares}, \hlc[pink]{some lattice surgery operations involve wider merged patches. This section investigates whether the window configuration in Figure 6 works on these patches, and the buffer width requirement for decoding on these patches.}
\begin{figure}
\centering
\sidesubfloat[]{
  \includegraphics[clip,scale=0.5]{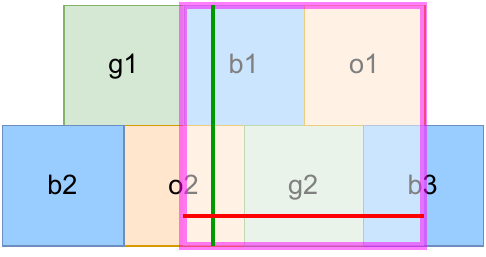}
  \label{subfig:top_right_config}
}\\
\vspace{2mm}
\sidesubfloat[]{
  \includegraphics[clip,scale=0.5]{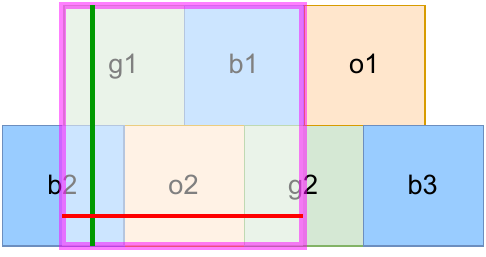}
  \label{subfig:top_left_config}
}
\caption{Setups for simulations with 5 windows stacked in 2 rows. As in \Cref{fig:staggered_squares}, the color of a window indicates the step that the decoder acts on it. The area of the underlying patch is enclosed by the pink square. The vertical and horizontal logical operators are in green and red. In (a) the two original patches that are merged are labelled b1 and o1, while in (b) they are labelled g1 and b1.}
\label{fig:2d_sim_configs}
\end{figure}

We perform simulations with the 2 settings shown in Figure 13, where we allocate an underlying patch in different regions of Figure 6, and perform decoding with the given window configurations. In practice, such a large patch could be the result of a merge: we start with two vertically-stacked square patches, each with the same size as a commit region in Figure 6. They are first deformed into rectangles with longer horizontal edges, so that they can perform $Y\otimes Y$ logical measurement~\cite{litinski2018lattice,chamberland2022circuit}. Then the top and bottom rectangles are merged. This example scenario motivates our choices of baselines: the total LERs of the two isolated patches, either before all the logical operations (when they are squares), or right before the merge (when they are longer rectangles).

\begin{figure}
\raggedright
\sidesubfloat[]{
  \includegraphics[clip,scale=0.8]{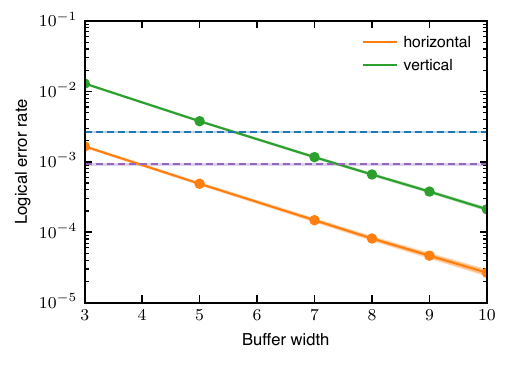}
  \label{subfig:top_right_results}
}\\

\sidesubfloat[]{
  \includegraphics[clip,scale=0.8]{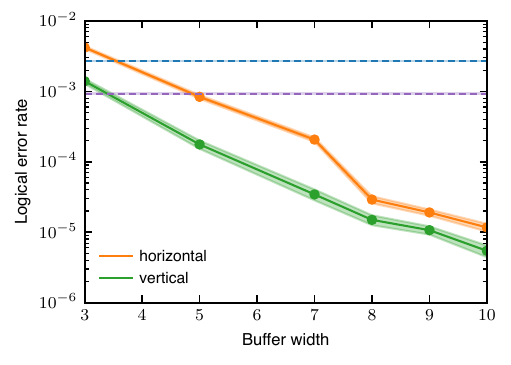}
  \label{subfig:top_left_results}
}
\caption{LER v.s. buffer width at $p=0.4\%$ over 15 rounds, using the setups (a) in Fig. 13a, and (b) in Fig. 13b. We show the different logical errors from using the horizontal and the vertical logical operators. The size of each smaller square (e.g. the commit region of b1) is chosen to be $15\times15$. The blue (purple) baseline is the LER of two isolated patches, each with size $15\times29$ ($15\times15$), taken over 15 rounds and with logical errors along the edge with $d=15$.}
\label{fig:2d_sim_results}
\end{figure}

\hlc[pink]{The simulation results in Figure 14 shows that, in this setting, the LER of parallel windows is suppressed exponentially with $w$, and is below the baseline before $w$ is grown to more than half the commit region width. Compared to the linear case presented earlier, the parallel windows can be seen to make use of the extra distance present during this logical operation --- the buffers extend in both directions rather than one direction. The difference between Figures 14a and b is due to the order in which the windows act. In (a), the case with vertical logical operator has higher errors because the intersection of the merged patch and the commit regions of o2 and b3 have short horizontal edges, and o2 and b3 are scheduled before their neighbor g2. This does not apply to (b) because in the lower row of (b), the middle window acts first.}

\subsection{Implications}\label{subsec:ler_implications}
The results in this section could help us determine the buffer width $w$ that is needed for achieving high accuracy. Linear settings (Figure 7) have larger buffer width requirements than multi-row settings (Figure 13), and the logical error rate in the linear setting is dominated by errors along the shorter edge. Therefore, it suffices to choose $w$ based on one's error budget and the physical noise level, according to the results in Section V-B. In general, $w$ should be around $d/2$. Its size could be larger if the physical noise level is low or if a closer match to the logical error rate of the global decoder is required. Conversely, it could be smaller if the physical noise level is high or if a looser match is acceptable.

\Cref{subsec:ler_long_edge} shows that the size of the windows (or specifically, the intersection of a patch with a window) also bound the fidelity that can be achieved. This is acceptable as long as a window is wider than an isolated patch of code. One mapping/compiling constraint should be noted though. When a patch spans multiple commit regions, and its intersection with one of the windows is too small, the type of logical errors in \Cref{fig:err_str_long} can be large. \hlc[pink]{This is the case in Figure 13a, where the intersection of the patch with the commit region of o2 has width $d/2$. When $w$ is at least half of $d$, o2's total width is at least $d$, so the contribution from this type of error is not too large. However, the intersection of a patch with a commit region should not be smaller than half the commit region.}

%% file: sections/throughput.tex
\section{Throughput}\label{sec:throughput}
In this section, we shift the focus to throughput \hlc[pink]{and study how large the window size can grow before the throughput limit is reached.} We quantify throughput by the average time that it takes for a decoder to process one round of measurements. Since each surface code measurement cycle takes $\sim 1~\mu s$ on a superconducting qubit-based quantum computer, we set it as the standard that a real time decoder should meet. 
\begin{figure}
  \centering
  \includegraphics[scale=0.55]{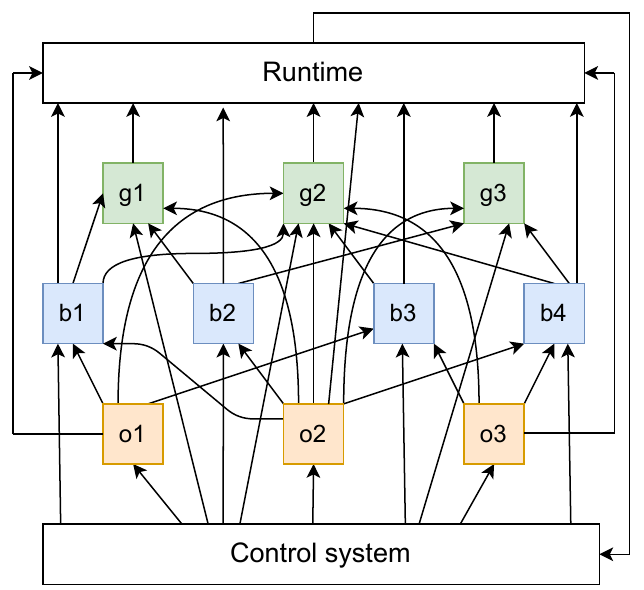}
  \caption{A flowchart of the parallel decoding scheme, the colored squares correspond to the windows in \Cref{fig:staggered_squares}.}
  \label{fig:flow_chart}
\end{figure}

\Cref{fig:flow_chart} shows a flowchart of a real time decoder using the spatially parallel windows. Each window receives information from the control system measuring the qubits and its neighboring windows from previous layers. After decoding, each window sends the committed corrections to the runtime and its artificial defects to neighboring windows in later layers.

\hlc[pink]{This results in a pipelined scheme} where the throughput is limited by the slower of two components: the inter-window communication or the inner decoder. \hlc[pink]{In this section, we first show that the inter-window communication will not be a bottleneck for an FPGA-based implementation, then discuss the scalability of inner decoders and how it affects the maximum window size that an FPGA implementation can have. To achieve larger window sizes than supported by FPGA-based inner decoders, one may turn to more resource efficient ASICs}~\cite{barber2023real}.
\subsection{Inter-window communication}\label{subsec:inter-window}
The throughput of the communication links depends on the number of bits that it takes to transmit information on the artificial defects.
As explained in \Cref{subsec:framework}, the only information that a decoder passes to its neighbor in a subsequent layer is the artificial defects. The length of this message is determined by (1) the number of bits required to represent an artificial defect and (2) the expected number of artificial defects that arise during decoding.

Quantity (1) depends on the number of different locations for artificial defects. Between two neighboring windows A and B, this is proportional to the area of $A_{\text{commit}}\cap B$, the intersection of window B and the commit region of A. 
Suppose we use the staggered square configuration, where the commit regions are $d\times d$ and $d$ cycles are processed at once. Then, between any pair of neighboring windows, the number of possible artificial defect positions is $\sim d^2/2$ for either the X or the Z decoding graphs. This translates to $\lceil \log_2(d^2/2)\rceil$ bits for representing each artificial defect — 9 bits for $d$ ranging from 23 to 32, and 12 bits for $d$ from 65 to 89.

\begin{figure}
  \centering
  \includegraphics[scale=0.8]{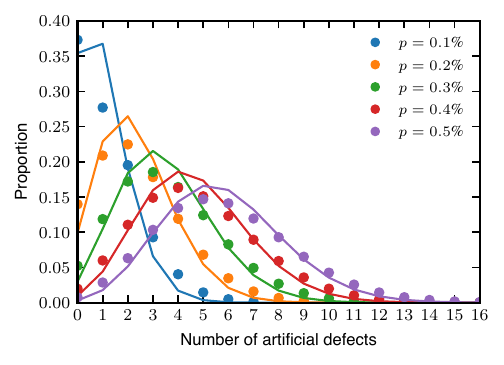}
  \caption{The number of artificial defects, from the same simulations as \Cref{fig:short_edge_ler}. The curves are fit to the Poisson distribution.}
  \label{fig:artificial_defects}
\end{figure}

To study quantity (2), we look at the distributions of artificial defects recorded from the simulations in \Cref{subsec:ler_short_edge} (\Cref{fig:artificial_defects}). We found that the distributions are independent of $w$ and heavily dependent on $p$. Distributions from different values of $p$ can be fit with the Poisson distribution with different means ($\lambda$), with the mean of the distribution roughly proportional to $p$. When we compared the data from \Cref{subsec:ler_short_edge} and \Cref{subsec:ler_long_edge}, we find that the mean of the distribution is also proportional to the area of $A_{\text{commit}}\cap B$. This is not surprising: $p$ is proportional to the chance that an artificial defect arises at each possible location, and the area of $A_{\text{commit}}\cap B$ is associated with the number of such locations.

Given $p$ and the area of the face where the artificial defects lie, we can estimate the mean count of artificial defects by extrapolating from our simulation results. For example, at $p=0.1\%$ and $d\sim 30$, the mean count of artificial defects is $\sim 5$. Then for each of the X and Z decoding graphs, it takes about $5\times 9 = 45$ bits to transmit the artificial defect information, without further optimization. This sums up to $\sim 90$ bits in total, for a link between two neighboring windows.

\hlc[pink]{For an FPGA-based implementation, each copy of the inner decoder is implemented on one FPGA and acts on one window. Therefore, the inter-window communication is between FPGAs.} For reference, the communication between two Xilinx FPGAs using a standard high-speed serial communication link is 54 clock cycles, plus 1 clock cycle per 64-bits~\cite{amd_manual}. On FPGAs, clock cycles of 4 ns (250 MHz) can be routinely achieved, even at high resource utilization. For windows with $d\times d$ commit regions arranged as in \Cref{fig:staggered_squares}, we have shown that, at $p=0.1\%$, the information passed between two neighbors is at most $\sim 94$ bits for $d\leq 30$, if $d$ rounds of syndrome is processed at once. That means each of these inter-FPGA links will take roughly 56 clock cycles, or $\sim 224$ ns if running at 250 MHz. Divided by $d$, the cost of communication is only 7 ns per measurement cycle, well below the requirement of 1 $\mu s$. We also calculated the communication overhead for larger $d$, and found that it decreases with $d$ in the range that is relevant. At $d=151$, the average cost of an inter-window communication is below 3 ns per measurement cycle. We performed the same calculation at a high physical noise of $p=0.5\%$, and found that the overhead is 44 ns per round at $d=5$ and steadily drops to below 9 ns at $d=151$. This is because when we divide the link latency by $d$, the contribution from the constant latency (54 clock cycles) decreases as $d$ increases, which outweighs the increase in the other term that is proportional to $d^2$. We conclude that the inter-window communication does not limit the size of windows.
\subsection{Inner decoders}\label{subsec:inner_decoder}
The throughput of the inner decoder depends on the size of the windows and the choice of decoder. Since spatially parallel windows can be layered over any inner decoder, and the development of real time decoders is still an active field, there is no clear answer for the maximum size of the window. Nonetheless, we can discuss the scalability of real time decoders that are currently available. Note that we also view the link from the control system to a window and the one from a window to the runtime as parts of an inner decoder.

\hlc[pink]{An inner decoder's scalability bottleneck depends on its design.} As examples, we look at two FPGA-based real time decoders \hlc[pink]{that work for at least medium-sized codes and have different scalability challenges}. Helios~\cite{liyanage2023scalable} is \hlc[pink]{a distributed version of a Union-Find decoder that exploits parallel computing resources for speedup. It has been demonstrated on a $21\times 21$ patch and operates with $O(d^3)$ processing elements, each assigned to a node in the decoding graph and communicates via shared memory.} Its average runtime per measurement round decreases with $d$, as long as $d$ is not large enough to require board to board communication between FPGAs \hlc[pink]{(which is much slower than shared memory based communication)}. Aside from this limit, the main scalability challenge of Helios is the $O(d^3)$ scaling of its computing resources. A recent update of Helios~\cite{godawatte2024multi} improved its scalability to a $51\times 51$ patch at the cost of increased latency.

\hlc[pink]{Riverlane's Collision Clustering (CC) decoder}~\cite{barber2023real} \hlc[pink]{is another UF-based decoder, with demonstrated implementations on both FPGA and ASIC. The advantage of CC is its efficient use of storage resources,} so its scalability is not constrained by hardware capacity, but instead constrained by the throughput. On an FPGA, its average execution time per measurement round increases with $d$, but is below $1~\mu s$ until $d=23$. \hlc[pink]{Like Helios, each instance of the CC decoder also needs to fit on 1 FPGA due to the usage of shared memory.}

The maximum window size supported by the FPGA is constrained by the scalability of the inner decoder. The decoding graphs in~\cite{barber2023real,liyanage2023scalable} are cubes, but for the spatially parallel windows, the number of measurement rounds that need to be processed at once does not depend on the full width of the window, but rather the width of the commit region. This would decrease the volume of the decoding graph by roughly half, but in practice, applying sliding windows along the time dimension would increase the volume of the decoding graph by $2X$. Therefore we expect that volume of the decoding graph in a window would be similar to the ones used in~\cite{barber2023real,liyanage2023scalable}, which means an inner decoder for $d=51$ translates to a commit region width of $\sim 25$ for the spatially parallel windows, assuming that the buffer is slightly wider than half of the commit region. We conclude that FPGA-based implementations of real time decoders connected into spatially parallel windows will be able to support at least medium term demonstrations of logical operations. In the long term, ASIC-based implementations should be able to support larger code distances, because they can be optimized for higher performance and lower cost~\cite{barber2023real}. 

\hlc{How does the maximum window size depend on the physical noise $p$?} Recall from \Cref{sec:ler} that when $p$ is lower, a wider buffer is required for the spatially parallel windows to achieve good accuracy. Wider buffers lead to larger windows, which impose higher requirements on the inner decoder. But it is worth noting that most decoders also run faster at low $p$~\cite{higgott2023sparse}, which is natural since physical errors are sparse at low $p$. If the main constraint on the scalability of an inner decoder is its throughput, as in the case of the CC decoder, then these are competing factors that determine the maximum window size it can support at low $p$.

%% file: sections/conclusion.tex
\section{Conclusion}
Spatially parallel windows make real time decoding of QEC codes scalable for the large patches of codes that arise during fault-tolerant quantum logical operations. We show that the scheme is compatible with the constraints imposed by using hardware accelerators, with proper window configuration. We perform numerical studies of the decoding accuracy and investigate the mechanisms through which the spatially parallel windows might introduce additional logical errors. The implications of the results impose constraints on mapping and introduce requirements on the window size and buffer width. We specifically study how the requirement on buffer width depends on the physical noise level. We assess the throughput of the decoding scheme and analyze the maximum window size that can be supported before the throughput falls below the requirement. We find that the communication between neighboring windows never becomes the bottleneck in the decoding scheme, so the maximum window size is limited by the scalability of the inner decoders. Since running on larger windows decreases the throughput of the inner decoders and/or requires more hardware resources, the window size and the buffer width need to be carefully chosen to achieve a balance between throughput/cost and accuracy. 